\documentclass[twocolumn,           % Format : preprint, twocolumn
               showpacs,            % Pacs : showpacs, noshowpacs
               nopreprintnumbers,     % Preprint: preprintnumbers,
               aps,                 % Society: ...
               prd,          	    % Journal Style : pra, prb, prc, prd, pre,
               letterpaper,             % Size : a4paper, ...
              groupeaddress,      % Affiliation (Title) : groupedaddress,
               nofootinbib,         % Footnote: footinbib, nofootinbib
               tightenlines,        % Remove additional spaces in a line
               floats,floatfix,      % Floating pictures and tables
               showkeys
               ]{revtex4-1}
               
\usepackage[toc,page]{appendix}
\usepackage{graphicx}% Include figure files
\usepackage{dcolumn}% Align table columns on decimal point
\usepackage{bm}% bold math
\usepackage{amsmath}
\usepackage{amsfonts,amssymb}
\usepackage{soul}
\usepackage{color}
\definecolor{v}{rgb}{0.6, 0.2, 0.8} %comentarios VM
\usepackage{xcolor}
\usepackage{enumerate}
\usepackage{float}
\usepackage{subfig}
\usepackage{ulem}

\begin{document}

\title{Cosmic acceleration in unimodular gravity}

\author{Miguel A. Garc\'ia-Aspeitia$^{1,2}$}
\email{aspeitia@fisica.uaz.edu.mx}

\author{C. Mart\'inez-Robles$^{1}$}
%\email{cesar.martinez@fisica.uaz.edu.mx}

\author{A. Hern\'andez-Almada$^3$}
%\email{ahalmada@uaq.mx}

\author{Juan Maga\~na$^{4,5,6}$}
%\email{juan.magana@uv.cl}

\author{V. Motta$^6$}
%\email{veronica.motta@uv.cl}

\affiliation{$^1$Unidad Acad\'emica de F\'isica, Universidad Aut\'onoma de Zacatecas, Calzada Solidaridad esquina con Paseo a la Bufa S/N C.P. 98060, Zacatecas, M\'exico.}
\affiliation{$^2$Consejo Nacional de Ciencia y Tecnolog\'ia, \\ Av. Insurgentes Sur 1582. Colonia Cr\'edito Constructor, Del. Benito Ju\'arez C.P. 03940, Ciudad de M\'exico, M\'exico.}
\affiliation{$^3$Facultad de Ingenier\'ia, Universidad Aut\'onoma de Quer\'etaro, Centro Universitario Cerro de las Campanas, 76010, Santiago de Quer\'etaro, M\'exico}
\affiliation{$^4$Instituto de Astrof\'isica, Pontificia Universidad Cat\'olica de Chile, Av. Vicu\~na Mackenna, 4860, Santiago, Chile.}
\affiliation{$^5$Centro de Astro-Ingenier\'ia, Pontificia Universidad Cat\'olica de Chile, Av. Vicu\~na Mackenna, 4860, Santiago, Chile.}
\affiliation{$^6$Instituto de F\'isica y Astronom\'ia, Facultad de Ciencias, Universidad de Valpara\'iso, Avda. Gran Breta\~na 1111, Valpara\'iso, Chile.}

%-------------------------------------------------------------------------------------------------
%-------------------------------------------------------------------------------------------------
\begin{abstract}
We study unimodular gravity in the context of cosmology, particularly some interesting consequences that might be able to describe the background cosmology and the late cosmic acceleration. We focus our attention on the hypothesis of \textit{non conservation of the energy momentum tensor}. This characteristic has an interesting outcome: we can obtain a modified Friedmann equation along with the acceleration equation and also new fluid equations related to a third order derivative of the scale factor, known in cosmography as the jerk parameter. As a consequence of this theory, it seems that radiation and the cosmological constant are intimately related, in agreement with what some authors have called the third coincidence problem. Their connection is the parameter $z_{ini}$, which has a value of $11.29$ and coincide with the reionization epoch. As a result, we are able to explain the late acceleration as a natural consequence of the equations, associating the new fluid to radiation and, thus, eliminating the need for another component (i.e. dark energy). Finally, we interpret the results and discuss the pros and cons of using the cosmological constant under the hypothesis of non conservation of the energy momentum tensor in the unimodular gravity scenario. 
\end{abstract}

%\draft
\date{\today}
\maketitle

%%%%%%%%%%%%%%%%%%%%%%%%%%%%%%%%%%%%
\section{Introduction}
%%%%%%%%%%%%%%%%%%%%%%%%%%%%%%%%%%%%
 Understanding the Universe acceleration at late times \cite{Riess:1998,Perlmutter:1999} is one of the most challenging puzzles in modern cosmology. This acceleration is attributed to an unknown entity called \textit{dark energy} (DE) under the Einstenian gravity,  but it could also be explained by modifying the theory of gravity. The first way implies that approximately sixty nine \cite{Planck:2018} percent of our universe is filled with something we do not know yet. Following this line of thought, some of the most studied models are: the cosmological constant (CC) \cite{Peebles:2003}, phantom and quintessence fluids \cite{Caldwell:2003vq,PhysRevD.37.3406}, generalized perfect fluids \cite{Hova2017,*Hernandez-Almada:2018osh}, etc.  The second possibility considers that the Einstein's theory of gravitation must be changed in order to explain this gravitational anomaly. A collection of different models are trying to understand how the theory of gravitation should be modified. For instance, the $f(R)$ theories \cite{Jaime:2012nj}, scalar-tensor theories \cite{Galiautdinov:2016qqy}, braneworlds  \cite{Aspeitia:2009bj,*Maartens:2010ar,*Garcia-Aspeitia:2016kak,*Garcia-Aspeitia:2018fvw,*Barcenas-Enriquez:2018ili}, Cardassian models \cite{FREESE20021,*Magana:2017}, among others. 

Even though there is a large number of models attempting to explain this phenomenon, the CC is the simplest and, according to the observations, apparently the preferred candidate responsible for the current Universe acceleration. This approach has been historically adopted in what is known as the $\Lambda$ cold dark matter (CDM) model, the standard model of cosmology.  Despite these positive attributes, $\Lambda$CDM exhibits some theoretical issues, e.g. the inability to match the observed value of the CC and the theoretical prediction coming from quantum field theory, where it is associated to the expectation value of the vacuum energy density \cite{Weinberg,*Zeldovich}. Another puzzling issue is the coincidence problem, i.e. why the Universe accelerates at $z\sim0.7$ and not after or before?

\textit{Unimodular gravity} (UG) \cite{Gao:2014nia,Josset:2016,Perez:2017krv,Perez:2018wlo} was first obtained by Einstein as an alternative field equation to GR\footnote{The vacuum solutions like Schwarzschild and Kerr remains unaltered in comparison with GR, the same happens for gravitational waves, provided that $\nabla^{\mu}T_{\mu\nu}=0$.} \cite{Einstein} (other references can be traced to \cite{James,VANDERBIJ1982307,Ellis}) and seems to be a serious alternative to face the problem of the Universe acceleration due to some interesting properties. First, in UG the determinant of the metric is kept fixed, instead of being a dynamical variable as in GR. This condition reduces the symmetry of the group of diffeomorphisms to the group of unimodular general coordinate transformations that leave the determinant of the metric unchanged. As a consequence, the new equations governing the dynamics of space-time are the trace-free Einstein equations, and now the vacuum energy has no direct gravitational effect. An important difference between them is the nature of the CC: while in GR it is a coupling constant in the Lagrangian, in UG it arises directly as an integration constant in the equations of motion. 

UG seems like a promising candidate to deal with the problems that afflict the CC, and it opens an interesting possibility of a natural violation of energy-momentum conservation (many authors propose the energy-momentum conservation as a separate assumption \cite{Ellis,Gao:2014nia}) that, although incompatible with GR,  it preserves coordinate transformations. This characteristic is the reason why UG was recently proposed by \cite{Josset:2016,Perez:2017krv,Perez:2018wlo} as a serious candidate to address the CC problem. The authors suggest that energy-momentum violations have been small but cumulative through the history of the Universe, those could be the cause of an effective CC with a value of the same order of magnitude that the one expected phenomenologically. Despite their interesting point of view, the authors in \cite{Perez:2017krv,Perez:2018wlo,Josset:2016} offer only a rather rough analysis of how the small violations of the energy conservation affect the standard evolution of the universe in their specific model.

In addition, it seems that preventing the vacuum energy from gravitating could be a possible path to avoid some of the CC issues. For instance, in \cite{Lombriser:2019jia} the author use this approach to calculate the density parameter of the CC, obtaining a constriction of $\Omega_{\Lambda}=0.704$.

In this paper we study some of the cosmological consequences of UG, maintaining \textit{the non conservation of the energy-momentum tensor}, and thus modifying not only the equations of the fluids but also the Friedmann and acceleration equations. The new generated terms manifest themselves in the background cosmology and undoubtedly also at the  perturbative level. The new correction terms in the dynamics involve the jerk parameter (JP) that depends on the third order derivative of the scale factor.  Choosing a particular JP, we allow the conservation of the equation of the fluids but maintain the new terms in the dynamical equations of UG. We interpret these new terms as some residual relativistic particles (radiation) that will be identified as the cause for the Universe late acceleration. This result has   profound implications for the existence of the CC and its relation with the hypothesis of the non conservation of the energy-momentum tensor.

We organize the paper as follows: in Sec. \ref{UG} we present the UG field equations, and obtain the movement equations in an homogeneous and isotropic cosmology, together with the non-conservation of the energy-momentum tensor. In Sec. \ref{Non} we show the modifications of Friedmann and acceleration equations in this scenario, presenting the master equations  involving the jerk parameter. In Sec. \ref{diagnostic} we reveal how a particular jerk recovers the traditional cosmology, introducing a new approach to resolve the problem of CC. Finally in Sec. \ref{CO} we give some conclusions and outlooks.

%%%%%%%%%%%%%%%%%%%%%%%%%%%%%%%%%%%%
\section{UG field equations} \label{UG}
%%%%%%%%%%%%%%%%%%%%%%%%%%%%%%%%%%%%

Unimodular gravity can be described by the Einstein-Hilbert action 
\begin{equation}
S = \int d^4x \xi \left(\frac{1}{16\pi G} R + \mathcal L_{matter}  \right)\, ,\label{action}
\end{equation}
where $\xi=\sqrt{-g}$, being $\xi$ a fixed scalar density which normally is called the unimodular condition and $\mathcal{L}_{matter}$ is associated with the matter density Lagrangian. 
It is important to remark that this condition is not the most formal way to define unimodular gravity: what it is really relevant
is that the equations of motion are obtained by considering an invariant volume form ($dV=\sqrt{-g}$ $dx^1\wedge dx^2\wedge...\wedge dx^4$). A volume form is coordinate independent,
while the unimodular condition is not. The physical consequence of the restricted variation considering $\xi=\sqrt{-g}$ in UG
is that the equations of motion are trace-free \cite{Ellis}. At the level of the equations of motion, we can impose any ansatz for the metric and locally rewrite them to satisfy the condition $\xi=\sqrt{-g}$. Therefore, it is irrelevant whether we start with a metric that satisfies the unimodular condition or not, at the end it is possible to obtain the same results with a coordinate transformation (see Appendix \ref{app} for the demonstration). Hence, after some calculations using \eqref{action}, we obtain the following field equations
\begin{equation}
R_{\mu\nu}-\frac{1}{4}g_{\mu\nu}R=8\pi G\left(T_{\mu\nu}-\frac{1}{4}g_{\mu\nu}T\right), \label{UGfield}
\end{equation}
where all the tensors are the standards of GR and $G$ is the Newton's gravitational constant. The previous equation can be rearranged in a more familiar way as
$G_{\mu\nu}+\Lambda(R,T)g_{\mu\nu}=8\pi GT_{\mu\nu}$,
where $\Lambda(R,T)\equiv\frac{1}{4}(R+8\pi GT)$ depends on the Ricci and energy-momentum scalars. Notice that, despite its resemblance, strictly speaking we do not recover the ten equations of GR, only nine independent equations.

These UG field equations could contain, in a natural way, an explanation for DE encoded in the term $\Lambda(R,T)$, which in general is not a constant function because it depends on the scalar functions $R$ and $T$. For a constant $\Lambda(R,T)\to\Lambda$ we return to the traditional interpretation of the CC and UG degenerates to general relativity as we will mention later. 

For a background cosmology, we consider an isotropic, homogeneous and flat Friedmann-Lemaitre-Robertson-Walker (FLRW) metric, $ds^2=-dt^2+a(t)^2d\vec{x}^2$, where $\sqrt{-g}=a(t)^3$ is not a constant function in general, however, through a coordinate transformation it is possible to obtain a metric that fulfill the unimodular condition. 
Notice that, if we start with the canonical FLRW metric, the physical interpretation is straightforward unlike the other case in which a coordinate transformation leads to $\xi=\sqrt{-g}$ (see Appendix \ref{app} for the demonstration of the equivalence between metrics). Therefore we will continue using the standard FLRW metric.

The perfect fluid energy momentum tensor is written as $T_{\mu\nu}=pg_{\mu\nu}+(\rho+p)u_{\mu}u_{\nu}$, where $p$, $\rho$ and $u_{\mu}$ are the pressure, density and cuadri-velocity of the fluid respectively. Hence, we have \cite{Ellis,Gao:2014nia}
\begin{equation}
\dot{H}=\frac{\ddot{a}}{a}-H^2=-4\pi G\sum_i(\rho_i+p_i). \label{Friedmann}
\end{equation}
The Hubble rate equation is composed by the traditional fluids except DE because it naturally emerges from UG.
In addition, we have the following equation:
\begin{equation}
8\pi G\nabla^{\mu}T_{\mu\nu}=\frac{1}{4}\nabla_{\nu}(R+8\pi GT)=\nabla_{\nu}\Lambda(R,T), \label{chida}
\end{equation}
which contains information about the conservation of the energy momentum tensor in the traditional GR form. Indeed, a general conservation for UG theory is now written in the form
\begin{equation}
\nabla^{\mu}[32\pi GT_{\mu\nu}-(R+8\pi GT)g_{\mu\nu}]=0. \label{CT}
\end{equation}
Hence, it is possible to infer the following possibilities. The first case is:

\begin{itemize}
\item assume the conservation of the energy-momentum tensor ($\nabla^{\mu}T_{\mu\nu}=0$) as an independent hypothesis, like in Ref. \cite{Gao:2014nia,Ellis}. The traditional cosmology is easily obtained with a CC as an integration constant, with no differences at the background level \cite{Gao:2014nia}. However, in the quantum realm, authors show important changes when compared to GR (see \cite{Saltas:2014cta,*Padilla:2014yea}). In this case, the CC does not suffer from a hierarchy problem and it is radiatively stable (see \cite{Smolin:2009ti,*Alvarez:2015sba,*Ardon:2017atk} for details).
\end{itemize}
The second case, which is the reason of our study,  
\begin{itemize}
\item assume the energy-momentum tensor is not conserved (see Eq. \eqref{chida}), without adding an extra assumption. This introduces new Friedmann, acceleration and fluid equations coupled with third order derivatives in the scale factor, which later we will relate to a cosmographic parameter. 
\end{itemize}

%%%%%%%%%%%%%%%%%%%%%%%%%%%%%%%%%%%%%%
\section{Non-conservation of the energy momentum tensor} \label{Non}
%%%%%%%%%%%%%%%%%%%%%%%%%%%%%%%%%%%%%%
If we assume the hypothesis of non conservation of the energy-momentum tensor, Eq. \eqref{chida}  must be solved to obtain the characteristic fluid equation. Solving for \eqref{chida} under a FLRW metric and perfect fluid we have
\begin{equation}
\sum_i\left[\frac{d}{dt}(\rho_i+p_i)+3H(\rho_i+p_i)\right]=\frac{H^3}{4\pi G}(1-j), \label{chida2}
\end{equation}
where the sum is over all the species in the Universe and $j\equiv\dddot{a}/aH^3$ is the Jerk Parameter (JP) \cite{Zhang:2016,Mamon:2018dxf}, well known in cosmography and this form is chosen to provide a more feasible interpretation. Notice that this theory is ruled by high order equations (third order derivatives to the scale factor and cubic exponents in the Hubble factor), implying a large number of initial conditions that could lead to spurious physical solutions, thus the need for a cosmographic quantity to provide physical interpretations. Consequently, we have given the JP a transcendental role as a guide to elucidate the behavior of Eq. \eqref{chida2} and its possible physical solutions.

The integral-transcendent-Friedmann equation can be computed with the help of Eq. \eqref{Friedmann} and \eqref{chida2}, obtaining 
\begin{equation}
H^2=\frac{8\pi G}{3}\sum_i\rho_i+H^2_{UG}(p_i,H,j,\mathcal{C}). \label{Frie}
\end{equation}
In addition, the acceleration equation is deduced from \eqref{Friedmann}, obtaining
\begin{equation}
\frac{\ddot{a}}{a}=-\frac{4\pi G}{3}\sum_i\left(\rho_i+3p_i\right)+H_{UG}^2(p_i,H,j, \mathcal{C}), \label{acc}
\end{equation}
where the non-canonical extra term in Eqs. \eqref{Frie}-\eqref{acc}, i.e. the UG contribution to the standard Friedmann and acceleration equations, is defined as
\begin{equation}
H_{UG}^2(p_i,H,j, \mathcal{C})\equiv\frac{8\pi G}{3}\sum_ip_i+\frac{2}{3}\int_{a_{ini}}^a H^2\left(j-1\right)\frac{da^{\prime}}{a^{\prime}} + \mathcal{C}, \label{HUG}   
\end{equation}
where the sum runs for the different species in the Universe, $a_{ini}$ is an initial value associated to the integral, and $\mathcal{C}$ is the integration constant. One of the main characteristics of UG is that it is possible to choose the value of the integration constant $\mathcal{C}$ to obtain a CC and hence a late Universe acceleration. However, the nature of the CC remains unknown as in $\Lambda$CDM. Therefore, we propose that there exist a protective symmetry that enforces $\mathcal{C}=0$ \cite{Hawking:1984,Coleman:1988,Perez:2017krv}, such that the source of the acceleration is the constant term associated with $a_{ini}$ in the integral. Hence, hereafter $H_{UG}(p_i,H,j,\mathcal{C})\to H_{UG}(p_i,H,j)$.

Another important feature is that the acceleration happens when $\int H^2(j-1)a^{-1}da>2\pi G(\rho+p)$ (see Eq. \eqref{acc}), implying that a fluid with negative EoS (strictly speaking $w<-1/3$) is not required, like in GR,  to accelerate the Universe.

%%%%%%%%%%%%%%%%%%%%%%%%%%%%%%%%%%%%%%
\subsection{Master equations in UG} \label{mastereq}
%%%%%%%%%%%%%%%%%%%%%%%%%%%%%%%%%%%%%%
We start this subsection assuming a barotropic fluid satisfying $w=p/\rho$, where $w$ is its constant equation of state (EoS). From the continuity equation \eqref{chida2}, we intuitively propose,
\begin{eqnarray}
\dot{\rho}_m+3H\rho_m=0, \;\;\to\;\;\rho_m=\rho_{0m}a^{-3}, \label{mat}
\end{eqnarray}
and
\begin{eqnarray}
\dot{\rho}_X+3H\rho_X=\frac{H^3}{4\pi G(1+w_X)}(1-j) \label{rad},
\end{eqnarray}
where only two main components are present: the matter (baryonic and DM) and the fluid coupled with the JP, hereafter called the X-fluid. 
We choose to separate Eq. \eqref{chida2}  into \eqref{mat} and \eqref{rad} to recover that matter evolves in the traditional form as $\rho_m\sim a^{-3}$, without extra hypothesis. The X-fluid's equation of continuity is coupled with the Hubble parameter and the JP. This coupling is one of the differences with the standard paradigm, remarking that the UG corrections are encoded in this equation.

Now, in order to study the UG, we propose the following dimensionless equations for the case of X-fluid and the Friedmann equation:  
\begin{eqnarray}
&&\frac{d\Omega_X(z)}{dz}-\frac{3}{(z+1)}\Omega_X(z)=\frac{2E(z)^2}{3(z+1)(w_X+1)}\times\nonumber\\&&[j(z)-1], \label{C2}\\
&&E(z)^2=\Omega_{0m}(z+1)^3+(1+w_X)\Omega_{X}(z)\nonumber\\&&-\frac{2}{3}\int_{z_{ini}}^z\frac{E(z)^2}{(z+1)}[j(z)-1]dz, \label{C3}
\end{eqnarray}
where we use the dimensionless definitions $E\equiv H/H_0$, $\Omega_i\equiv8\pi G\rho_i/3H_0^2$, and the matter equation is already solved in \eqref{mat} and added into the previous equations. We also notice that the value $w_X=-1$ is forbidden from Eq. \eqref{C2}, whose value coincide with the EoS for the cosmological constant in GR. With regard to $z_{ini}$, it will play a preponderant role in the following calculations because it will generate a non gravitational constant that hereafter will be interpreted as the cause of the Universe acceleration. 

In addition, for the deceleration parameter we write
\begin{eqnarray}
q(z)&=&\frac{1}{2E(z)^2}\Big\lbrace\Omega_{0m}(z+1)^3+(1+w_X)\Omega_X(z)\nonumber\\&& +\frac{4}{3}\int_{z_{ini}}^z E(z^{\prime})^2[j(z^{\prime})-1]\frac{dz^{\prime}}{(z^{\prime}+1)}\Big\rbrace, \label{qN}
\end{eqnarray}
here $z_{ini}$ is the same initial value related to $a_{ini}$. 
Hereafter, we will call Eqs. \eqref{C2}-\eqref{C3} the master equations for any analysis in UG.

%%%%%%%%%%%%%%%%%%%%%%%%%%%%%%%%%%
\section{JP Diagnostic and Results} \label{diagnostic}
%%%%%%%%%%%%%%%%%%%%%%%%%%%%%%%%%%%%%%
Section \ref{mastereq} shows that the JP plays an important role in the master equations of the UG, however the UG does not give information of its characteristics or its functional form. This is significant because it is related to the non conservation of the energy-momentum tensor and could help us to choose an ansatz. Hence, many forms of the JP can be proposed to model the observed Universe. Nevertheless, we expect a JP with the capability of reproducing the observed cosmology, i.e it should not affect the structure formation nor Nucleosynthesis, to maintain the well established knowledge. 

One interesting way is demanding the conservation of the X-fluid's continuity equation through a given JP. This is achieved with the equation
\begin{equation}
j(z)=\frac{9(1+w_X)w_X}{2E(z)^2}\Omega_{0X}(z+1)^{3(1+w_X)}+1, \label{JX}
\end{equation}
that easily helps to integrate the master equations resulting in
\begin{eqnarray}
E(z)^2&=&\Omega_{0m}(z+1)^3+\Omega_{0X}(z+1)^{3(w_X+1)}\nonumber\\&&+w_X\Omega_{0X}(z_{ini}+1)^{3(w_X+1)}, \label{HX} 
\end{eqnarray}
together with the deceleration parameter as
\begin{eqnarray}
q(z)&=&\frac{1}{2E(z)^2}\Big[\Omega_{0m}(1+z)^3+(1+3w_X)\Omega_{0X}(z+1)^{3(w_X+1)}\nonumber\\&&-2w_X\Omega_{0X}(z_{ini}+1)^{3(w_X+1)}\Big], \label{qX}
\end{eqnarray}
where $w_X$ must be positive to obtain a d'Sitter behavior. Notice that new physic terms arise (in comparison with the standard model) in the $H(z)$ and $q(z)$ parameters, associated to a constant that depends on the characteristics of the X-fluid and its equation of state. Hence, the only causative of the acceleration is:
\begin{equation}
\Omega_{0\Lambda}=w_X\Omega_{0X}(z_{ini}+1)^{3(w_X+1)}, \label{OLam}
\end{equation}
which is an integration constant that does not gravitate, relieving the afflictions contained in GR.

\begin{figure}
\centering
\par\smallskip
{\includegraphics[width=0.5\textwidth]{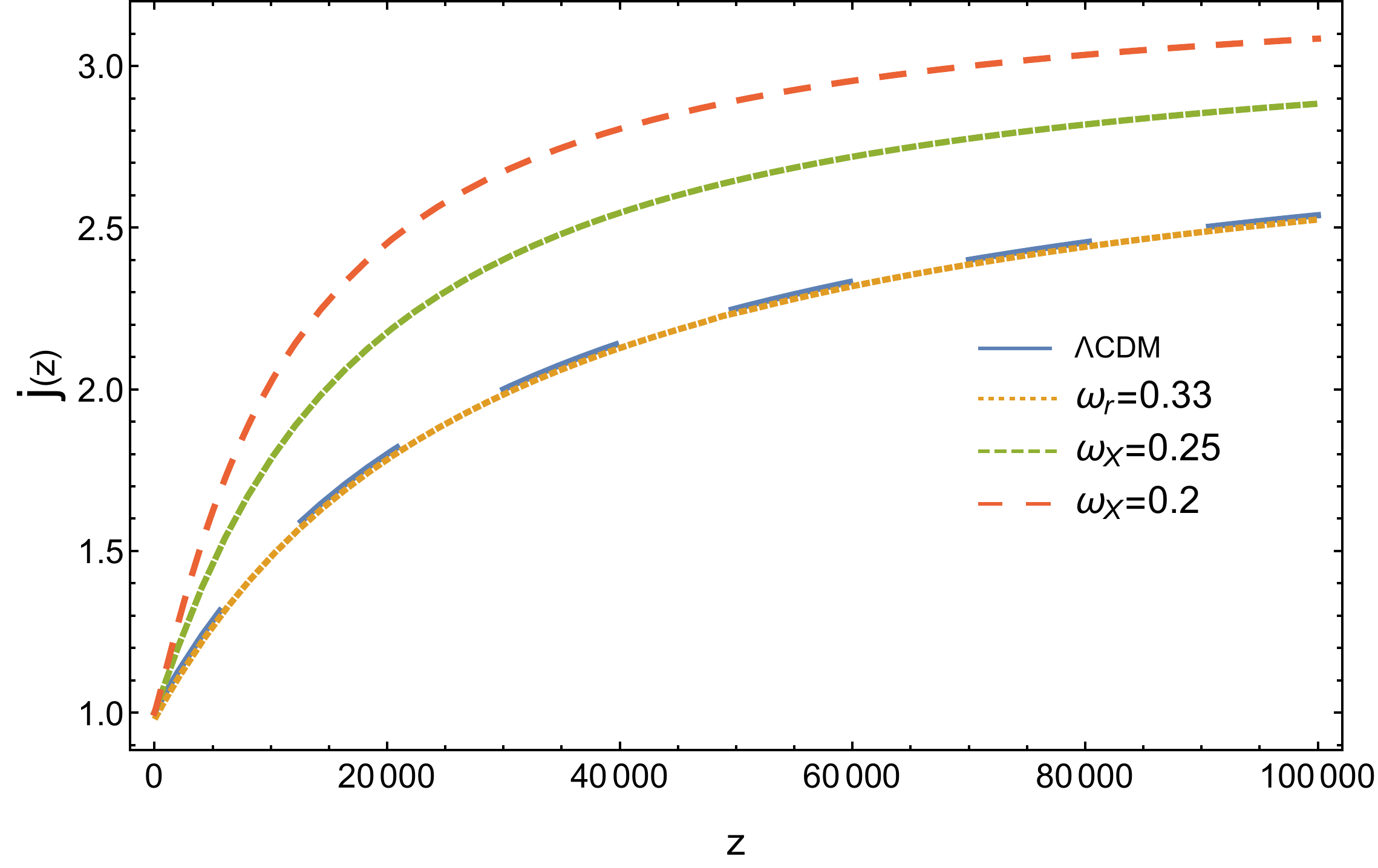}}
\caption{Jerk parameter for UG and $\Lambda$CDM, assuming $\Omega_{0r}=9.07\times10^{-5}$ for both theories and $w_r=0.33$, $w_X=0.25$ and $w_X=0.2$. We observe that UG coincide with $\Lambda$CDM-like only when $w_X=w_r=0.33$, as expected. JP for $\Lambda$CDM-like is computed through the formula $j(q)=q(2q+1)+(1+z)dq/dz$ \cite{Mamon:2018dxf}, assuming also $\Omega_{0m}=0.31$ and with that $\Omega_{0\Lambda}=0.69$ \cite{Planck:2018}.}
\label{Figure:jerk}
\end{figure}

This posses the question: to what kind of fluid does $\Omega_X(z)$ corresponds to? The first idea is to make a correspondence with radiation (which is the missing fluid in our equations, in addition to the DE, and represents the relativistic particles), i.e. $\Omega_X\to\Omega_r$, being $w_X=w_r=1/3$. 
In this way, the standard $\Lambda$CDM is mimicked with the constant $\frac{1}{3}\Omega_{0r}(z_{ini}+1)^4$, which is the cause of the acceleration and the new interpretation of the CC. Fig. \eqref{Figure:jerk} shows several JP given by Eq. \eqref{JX} with different EoS. Notice that the best candidate, as expected, is the case $w_r=1/3$, producing the evolution not only in recent epochs but also in the early Universe.

One main concern is the value for $\Omega_{0r}$, which is constrained as $2.469\times10^{-5}h^{-2}(1+0.2271g_{*})$, where $g_{*}=3.04$ is the standard number of relativistic species \cite{Komatsu:2011} and $h=0.678$ is the dimensionless Hubble constant according to Planck satellite observations \cite{Planck:2018}. In this case, $z_{ini}$ is an important factor that relates the $\Omega_{0\Lambda}$ value with $\Omega_{0r}$.  We obtain the observed density parameter of $\Omega_{0\Lambda}=0.69$ for $z_{ini}\approx11.29$ as a lower bound, equivalent to $\sim 13.17$~Gyrs in the past, which coincide with the reionization epoch. Thus, the relation $\Omega_{0\Lambda}\equiv\frac{1}{3}\Omega_{0r}(z_{ini}+1)^4$ found in our analysis shed more light on the source of such relation. 

On the other hand, recently some authors have claimed, based on a statistical comprehensive study of observations, that there might be another coincidence problem at the epoch of reionization. According to Ref. \cite{Lombriser:2017cjy}, the observations suggest that around  $z=9.6$ the energy density of radiation and the energy density of the CC coincide. In our case, the temperature at which the Universe should have an accelerated stage can be estimated as follows. The relation between the density parameters gives $
\rho_{0\Lambda}=\frac{1}{3}\rho_{rad}(z_{ini}+1)^4$,
where we identify $\rho_{rad}=\pi^2g_*T^4/30$ \cite{peter2009primordial} and $T$ the temperature predicted by the model, hence we have the equation
\begin{equation}
    T=\left(\frac{90\rho_{0\Lambda}}{\pi^2g_*}\right)^{1/4}(z_{ini}+1)^{-1}. \label{Tacc}
\end{equation}
Substituting $z_{ini}=11.29$, $\rho_{0\Lambda}\leqslant(10^{-12}\rm{GeV})^4$, and $g_*=3.04$ we obtain a temperature of $\sim1.07\times10^{-4}$eV similar to $T_{0CMB}\approx2.35\times10^{-4}$eV; what lead us to think that Eq. \eqref{Tacc} suggests a possible path to resolve the problem of coincidence.

To finish this section, we discuss the possibility of other JP's  mimicking the standard cosmological behavior but using Eq. \eqref{C2} with a different evolution, i.e. without resembling radiation.
Therefore, we assume that, at late times, Eq. \eqref{C2} transmute from a radiation component to a new $X$-fluid under the change of the jerk and EoS parameters. This premise is the cause of the aforementioned subtle differences with the standard model. For example, sinusoidal jerk functions with negative EoS for the $X$-fluid can mimic the standard model, yielding testable differences in the $H(z)$ and deceleration parameter. It is possible to explore a  sinusoidal jerk like $j(z)=1+j_0\sin^n(z)+j_1\cos^n(z)$,
where $j_i$ ($i=0,1$) and $n$ odd are free parameters adjusted in order to obtain a behavior comparable to the standard cosmological model \cite{Tamayo:2019gqj}. In this case, it is not clear which value for EoS must be assigned and also it is not possible to recover the presence of radiation. In addition, the sinusoidal jerk only fits the epoch of matter and CC, leaving aside the radiation epoch in contrast to what we found in Eq. \eqref{JX}. 

%%%%%%%%%%%%%%%%%%%%%%%%%%%%%%%%%%%%%%
\section{Conclusions and Outlooks} \label{CO}
%%%%%%%%%%%%%%%%%%%%%%%%%%%%%%%%%%%%%%
UG was studied for decades but its interest as a plausible explanation for the CC was renewed by authors in \cite{Perez:2017krv,Perez:2018wlo,Josset:2016}. In this vein, we explore UG in a cosmological context, investigating its viability to reproduce the current Universe acceleration, improving and confirming previous findings. 

If we assume that the energy momentum tensor is not conserved through Eq. \eqref{chida}, the implications to the continuity equation and, in consequence, to the Friedmann equation are important in comparison to the standard paradigm; both new equations contain a third derivative in the scale factor that is interpreted as the JP. It is worth noting that the way the equation of continuity is written suggests that the relativistic particles (radiation and neutrinos) must be coupled with the JP, while matter follows the standard behavior.
Choosing an appropriate JP that conserve the continuity equation of the fluid and relates to relativistic particles behavior implies important consequences. The first one is that UG can reproduce the standard cosmology with two constants that emerge naturally from the theory and, therefore, produces a late Universe acceleration. Moreover, one of those constants has information about the fluid with the form $\frac{1}{3}\Omega_{0r}(z_{ini}+1)^4$. Regarding the other constant, a protective symmetry suggests that it must be zero, implying that only the first term is the cause of the acceleration. Under these premises and the comparison with observations, we conclude that $z_{ini}\approx11.29$ in order to agree with the density parameter of CC, which suggestively comes from reionization epoch\footnote{Redshift bounds for the reionization epoch can be considered as $6\lesssim z\lesssim20$, however, the reader can examine other references with more stringent bounds for reionization in \cite{Aghanim:2018,*Zaroubi,*Barkana:2000}.} and points towards the existence of a more fundamental relation between radiation and the CC. In this vein, we suggest the possibility that future observations of Lyman-$\alpha$ emissions \cite{Fan:2005,*Tilvi:2014,*Schenker:2014t} could be the key to refute or validate the scenario propounded in this letter.
In addition, these results generate another important consequence: the temperature predicted by this model suggestively coincide with the current CMB temperature. Thus, we interpret this result as being the temperature where the Universe should begin its acceleration epoch, providing a possible explanation for the coincidence problem. 

 In essence, our thesis is that this new physics is governed by and encoded in the JP form, although the origin of this JP and its deduction through first principles is not clear yet\footnote{However, this is also a problem affecting the CC.}, at least from the UG point of view. Other more fundamental theories could give us clues about the origin of this JP. 

Finally, the most outstanding feature of UG is the possibility of solving the Universe acceleration problem by adding a function that resembles the CC and the standard cosmology. This function could be associated to radiation, avoiding the need to assume an unknown fluid (i.e. dark energy) to explain the late expansion. Another strong point of this theory is the potential evidence of the energy conservation violation associated with the no conservation of the energy-momentum tensor, which several authors interpret as possible corroboration of the space-time granularity. 
%However this is beyond the scope of this paper.

\begin{acknowledgements}
%%%%%%%%%%%%%%%%%%%%%%%%%%%%%%%%%%%%%%%%%%%%%%%
 C.M.-R. acknowledges the support provided by CONACyT Ph.D. fellowship;
M.A.G.-A. acknowledges support from SNI-M\'exico, CONACyT research fellow, COZCyT and Instituto Avanzado de Cosmolog\'ia (IAC) collaborations. We thank the enlightening conversations and valuable feedback with Daniel Sudarsky, Alejandro P\'erez, Michel Cur\'e, Luis Ure\~na, Javier Chagoya, Yasm\'in Alc\'antara and Leonardo Quintanar. Also M.A.G.-A. acknowledges the hospitality of the Instituto de
F\'isica y Astronom\'ia of Universidad de Valpara\'iso,
Chile, where part of this work was done. J.M. acknowledges the support from Fondecyt 3160674 and CONICYT project Basal AFB-170002, V.M. acknowledges the support of Centro de Astrof\'{\i}sica de Valpara\'{\i}so (CAV). \\
\end{acknowledgements}

\appendix
%%%%%%%%%%%%%%%%%%%%%%%%%%%%%%%%%%%
\section{Demonstration of Equivalence} \label{app}
%%%%%%%%%%%%%%%%%%%%%%%%%%%%%%%%%%%
Here we show the equivalence between the FLRW metric and another one that fulfills the unimodular condition. According to Alvarez et al. \cite{Alvarez:2015sba}, the cosmology in UG can be treated with the metric
\begin{equation}
    ds^2=-b(\tau)^{-2/3}d\tau^2+b(\tau)^{1/2}d\vec{x}^2, \label{ucondition}
\end{equation}
which is written in unimodular coordinates.
Notice that Eq. \eqref{ucondition} can be constructed, using standard FLRW line element and the change of variables $a\to b^{1/4}$ and $dt\to b^{-3/4}d\tau$.

Assuming a perfect fluid energy-momentum tensor and using UG equations (see Eq. \eqref{UGfield}) we have
\begin{equation}
\frac{b^{\prime\prime}}{b}-\frac{1}{4}\left(\frac{b^{\prime}}{b}\right)^2=-16\pi Gb^{-3/2}(\rho+p),   
\end{equation}
where primes indicate derivatives with respect to $\tau$. This is the Friedmann equation under the unimodular condition, but the physical interpretation in this form is not straightforward. However, if we return to the $a(t)$ and $t$ variables to recover the traditional FLRW line element, we finally have: $\dot{H}=-4\pi G(\rho+p)$,
which is the same equation that we use in our previous analysis. Therefore, at least in this case, the result is independent of using the FLRW metric or the metric of Eq. \eqref{ucondition}. The essential point is to choose the metric that provides the best insight into the physical interpretation of the results.
%-------------------------------------------------------------------------------------------------
%------------------------------------------------------------------------------------------------
\bibliography{librero1}

\end{document}